\documentclass[manuscript]{acmart}
\usepackage{amsmath}

\usepackage{amsfonts}
\usepackage{algorithmic}
\usepackage{array}
\usepackage[caption=false,font=normalsize,labelfont=sf,textfont=sf]{subfig}
\usepackage{textcomp}
\usepackage{stfloats}
\usepackage{url}
\usepackage{verbatim}
\usepackage{graphicx}
\usepackage{xspace}

\usepackage{multirow}
\def\BibTeX{{\rm B\kern-.05em{\sc i\kern-.025em b}\kern-.08em
    T\kern-.1667em\lower.7ex\hbox{E}\kern-.125emX}}
\usepackage{balance}

\def \ourtool{WACANA\xspace}

\begin{document}

\title{\ourtool: A Concolic Analyzer for Detecting On-chain Data Vulnerabilities in WASM Smart Contracts}

\author{Wansen Wang}
\email{23762@ahu.edu.cn}
\affiliation{%
  \institution{Anhui University}
  \city{Hefei}
  \state{Anhui}
  \country{China}
}

\author{Caichang Tu}
\email{tucaichang@mail.ustc.edu.cn}
\affiliation{%
  \institution{University of Science and Technology of China}
  \city{Hefei}
  \state{Anhui}
  \country{China}
}

\author{Zhaoyi Meng}
\authornote{Corresponding authors: Zhaoyi Meng, Wenchao Huang}
\email{22112@ahu.edu.cn}
\affiliation{%
  \institution{Anhui University}
  \city{Hefei}
  \state{Anhui}
  \country{China}
}

\author{Wenchao Huang}
\authornotemark[1]
\email{huangwc@ustc.edu.cn}
\author{Yan Xiong}
\email{yxiong@ustc.edu.cn}
\affiliation{%
  \institution{University of Science and Technology of China}
  \city{Hefei}
  \state{Anhui}
  \country{China}
}

\begin{abstract}
WebAssembly (WASM) has emerged as a crucial technology in smart contract development for several blockchain platforms.
Unfortunately, since their introduction, WASM smart contracts have been subject to several security incidents caused by contract vulnerabilities, resulting in substantial economic losses.
However, existing tools for detecting WASM contract vulnerabilities have accuracy limitations, one of the main reasons being the coarse-grained emulation of the on-chain data APIs.
In this paper, we introduce \ourtool, an analyzer for WASM contracts that accurately detects vulnerabilities through fine-grained emulation of on-chain data APIs. \ourtool precisely simulates both the structure of on-chain data tables and their corresponding API functions, and integrates concrete and symbolic execution within a coverage-guided loop to balance accuracy and efficiency. Evaluations on a vulnerability dataset of 133 contracts show \ourtool outperforming state-of-the-art tools in accuracy. Further validation on 5,602 real-world contracts confirms \ourtool's practical effectiveness. 
\end{abstract}



\maketitle

\section{Introduction}

In recent years, WebAssembly (WASM)~\cite{wasm} has emerged as a crucial technology in smart contract development for several blockchain platforms, including EOSIO~\cite{eosio} and NEAR~\cite{near}. Additionally, a development team is working on implementing an upgraded Ethereum virtual machine (VM), which aims to replace the existing EVM with a WASM-compatible VM~\cite{ewasm}. 

Unfortunately, since their introduction, WASM smart contracts have been subject to many security incidents caused by contract vulnerabilities, resulting in substantial financial losses.
The EOSIO platform, being the first to support WASM contracts, provides several examples of such security breaches. 
In 2018, EOSBet, a gambling contract on the EOSIO platform, lost approximately \$540,000 in EOS due to two attacks exploiting the Fake EOS vulnerability and the Fake Notification vulnerability~\cite{eosattack1}\cite{eosattack2}. 
Subsequently, in 2021, the flash.sx smart contract fell victim to a reentry attack, resulting in the theft of about 1.2M EOS (approximately \$13M )~\cite{eosattack3}. 
More recently, in 2023, the EOS project pcash suffered an attack leading to losses of around \$2 million~\cite{eosattack4}. 
These incidents highlight the critical need for effective methods to detect and prevent vulnerabilities in WASM smart contracts.

Current research on vulnerability detection for WASM smart contracts primarily follows two kinds of approaches:
\begin{itemize}
\item The first kind of approach is based on fuzzing test, exemplified by tools like EOSFUZZER~\cite{huang2020eosfuzzer} and WASAI~\cite{chen2022wasai}. EOSFUZZER generates fuzz inputs based on contract interface information, executes the contract using a modified WASM virtual machine, and collects data for vulnerability detection. 
WASAI enhances this approach by performing symbolic simulations on execution traces to construct path constraints, thereby improving EOSFUZZER's random seed generation strategy. 
However, these tools often struggle with achieving comprehensive code coverage due to the inherent limitations of fuzzing test.

\item The second kind of approach involves symbolic execution-based methods, which typically achieve higher code coverage by traversing all possible paths through symbolized inputs rather than relying on finite test cases~\cite{pak2012hybrid}. 
Notable examples include EOSAFE~\cite{he2021eosafe}, which performs symbolic execution on WASM bytecodes and employs vulnerability-specific heuristic pruning to address state explosion issues, while using on-demand emulation of library functions. 
WANA~\cite{jiang2021wana} provides comprehensive support for the WASM specification 1.0 instruction set~\cite{spec1} and includes emulation of Ethereum platform library functions, making it adaptable to Ethereum WASM smart contracts. 
\end{itemize}

While symbolic execution-based approaches have become a more commonly used one due to their higher code coverage, these approaches share a common limitation: the inaccurate modeling of on-chain data APIs. While EOSIO has introduced an on-chain data mechanism to provide persistent data storage for WASM smart contracts~\cite{datapersist}, which has been widely adopted across numerous WASM contracts, current analyzers employ only coarse-grained emulation of on-chain data API functions. For instance, WANA merely uses random values to represent on-chain data without emulating the actual API function steps, while EOSAFE simply records the function names and arguments rather than analyzing the internal operations. These abstraction-based approaches potentially compromise the accuracy of vulnerability detection, particularly when dealing with vulnerabilities related to on-chain data APIs.

In this paper, we introduce \ourtool, an analyzer for WASM contracts that accurately detects vulnerabilities through fine-grained emulation of on-chain data APIs. Our tool precisely simulates both the structure of on-chain data tables and their corresponding API functions, enabling detailed modeling of operations such as add, delete, modify, and retrieve during symbolic execution. To balance accuracy and efficiency while reducing the time overhead from this fine-grained API emulation, we employ a hybrid approach that integrates concrete and symbolic execution within a coverage-guided loop:
% To address the challenge of state explosion resulting from symbolic simulation of on-chain data tables, we 

\begin{enumerate}
\item Deploy a test EOS chain and intialize an empty on-chain database.
\item Execute each function of contracts symbolically and detect predfined vulnerability patterns while tracking the resulting state; 
\item Compare instruction coverage of each resulting state against the initial state; 
\item When increased coverage is detected, identify the state with maximum coverage, concretize its corresponding function parameters in a transaction and execute it on the test chain to obtain the new on-chain database, and repeat from step 2; 
\item Terminate execution and report detected vulnerabilities when no further coverage increases are found. 
\end{enumerate}

% This goal of this coverage-guided mechanism  is to balance the analysis complexity and code coverage.

What's more, we demonstrate \ourtool's effectiveness through two evaluations. 
First, we conduct a comparative analysis against WANA and WASAI using a dataset of 133 smart contracts with known vulnerabilities. 
The results show our tool's superior performance across key metrics including precision, recall, and F1 value. 
Second, we validate our tool's practical applicability by testing it against a real-world dataset of 5,602 contracts, confirming its capability to detect vulnerabilities in real-world environments.

In summary, the contributions of this paper are:
\begin{itemize}

\item  We propose a fine-grained emulation method for on-chain data APIs of WASM contracts, providing more accurate modeling of data structures and dependencies compared to existing coarse-grained approaches.

\item We propose a coverage-guided concolic execution algorithm that leverages concrete execution to reduce symbolic execution paths, reducing the analysis complexity while maintaining accuracy.

\item Based on the above methods, we implement \ourtool, a concolic analyzer for WASM smart contracts that enhances the detection accuracy of vulnerabilities related to on-chain data while optimizing analysis efficiency through integration of concrete execution.

\item We compare the effectiveness of \ourtool with other representative tools on the vulnerability dataset consisting of 133 contracts. The experimental results show that our tool outperforms other tools in terms of precision, recall, and other metrics. In addition, we experimentally demonstrate the effectiveness of components, such as emulation of on-chain data APIs and concolic execution, on this dataset.

\item We perform an evaluation on 5,602 real-world contracts, with detailed manual analysis of a 200-contract subset confirming a 97.09\% recall rate, demonstrating \ourtool's practical effectiveness.
\end{itemize}

\section{Background}
\label{sec:back}

\begin{figure*}
\centering
\includegraphics[scale=0.44]{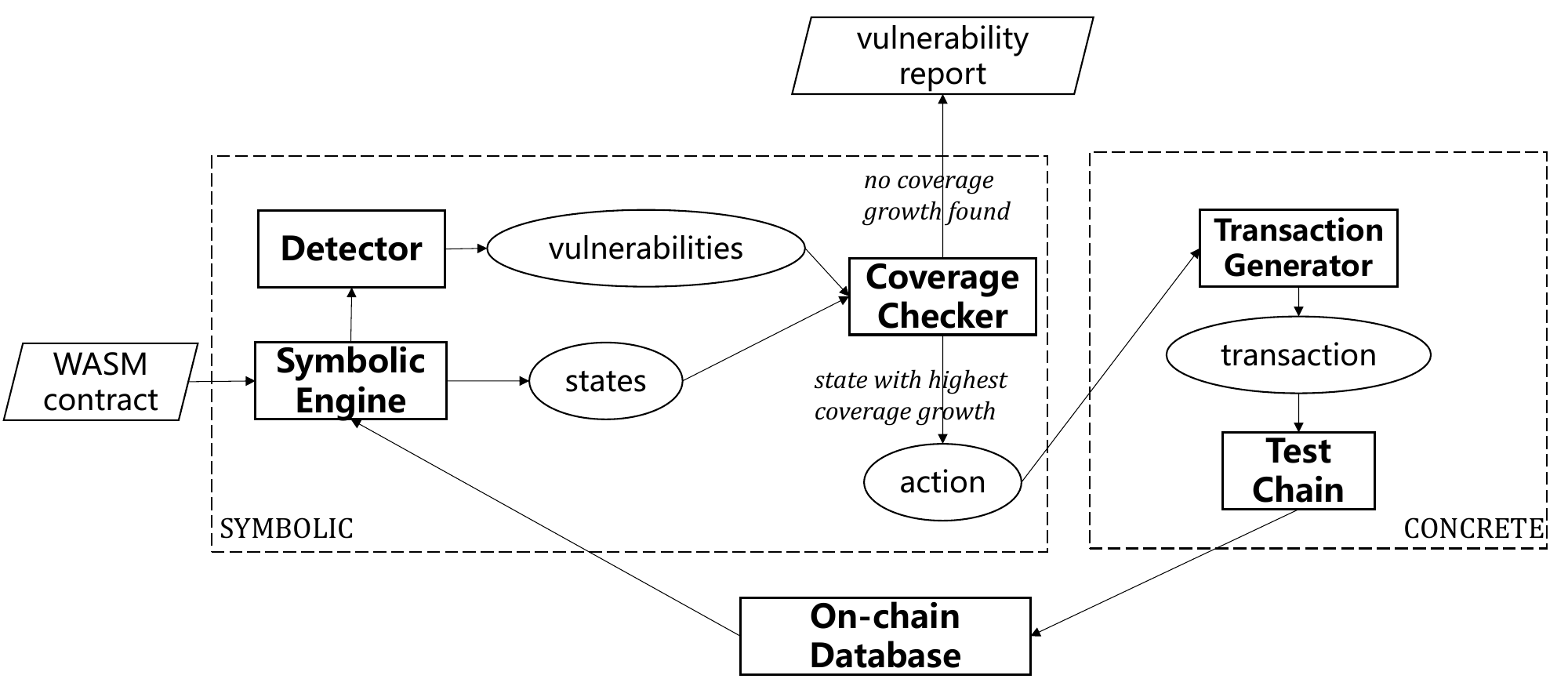} 
\caption{Overview of \ourtool.}
\label{fig:design}
\end{figure*}

In this paper, we focus on EOSIO, one of the most popular platforms currently supporting WebAssembly (WASM) smart contracts. This section introduces the fundamental concepts of EOSIO and its WASM-based smart contract system.

\textbf{Accounts, smart Contracts and actions.} On the EOSIO platform, participants are identified by accounts, which are associated with smart contracts implemented in WASM code.
Accounts can interact with each other through actions, and the associated WASM code processes incoming actions by calling corresponding functions.
Actions define atomic behaviors within these smart contracts, which can be combined into transactions representing the smallest unit for validation and blockchain integration. 
While actions are typically generated through user input, they can also be created dynamically by contracts at runtime through inline actions, using serialized action data with the $\textsl{send\_inline}$ function.

\textbf{VM Data Storage and On-chain Database.}
The EOS virtual machine~\cite{eosvm}, which is used to execute WASM contracts, employs two distinct categories of data storage areas during contract execution. The first category is function-scoped storage, which includes the $\textsl{Stack}$ area for temporary instruction execution values and the $\textsl{Local}$ area for function parameters and local variables. The second category consists of cross-function storage areas: the $\textsl{Global}$ area for simple global variables and the $\textsl{Memory}$ area for complex data structures manipulated through $\textsl{load}$ and $\textsl{store}$ operations. 

While these virtual machine storage areas are reinitialized with each execution, persistent data storage is achieved through the on-chain database mechanism. This mechanism enables developers to create custom database structures with comprehensive CRUD (Create, Read, Update, Delete) API support. The on-chain database is organized into tables, each uniquely identified by a combination of owner, scope, and table name. Within these tables, data is stored as key-value pairs, with each row accessible through an iterator (functioning similarly to a pointer). Data retrieval follows a specific sequence: contracts must first locate the target table, obtain the appropriate iterator, and then access the desired data row using this iterator.

\textbf{Existing vulnerabilities.} According to \cite{vulexample}, there are several kinds of vulnerabilities in the EOSIO smart contracts that are widely studied:
\begin{itemize}
\item \textit{Blockchain-info Dependency.}
Smart contracts, particularly those implementing gambling features, require secure random number generation algorithms to ensure game fairness in applications such as betting systems. Many such contracts attempt to achieve randomness by utilizing blockchain runtime information (like block numbers) as seeds through library functions. However, due to the public nature of contract binaries on EOSIO platform, attackers can reverse engineer these binaries to understand the random number generation algorithm, and since the blockchain runtime information used as seeds is often predictable or potentially controllable, they can strategically time their bets to exploit the system and compromise game fairness.

\item \textit{Rollback.}
When an inline action in a transaction fails, the platform automatically rolls back the entire transaction, including all associated actions. This feature, while designed for transaction integrity, can be exploited by malicious actors to selectively cancel unfavorable outcomes.
\begin{figure}
\centering
\includegraphics[scale=0.33]{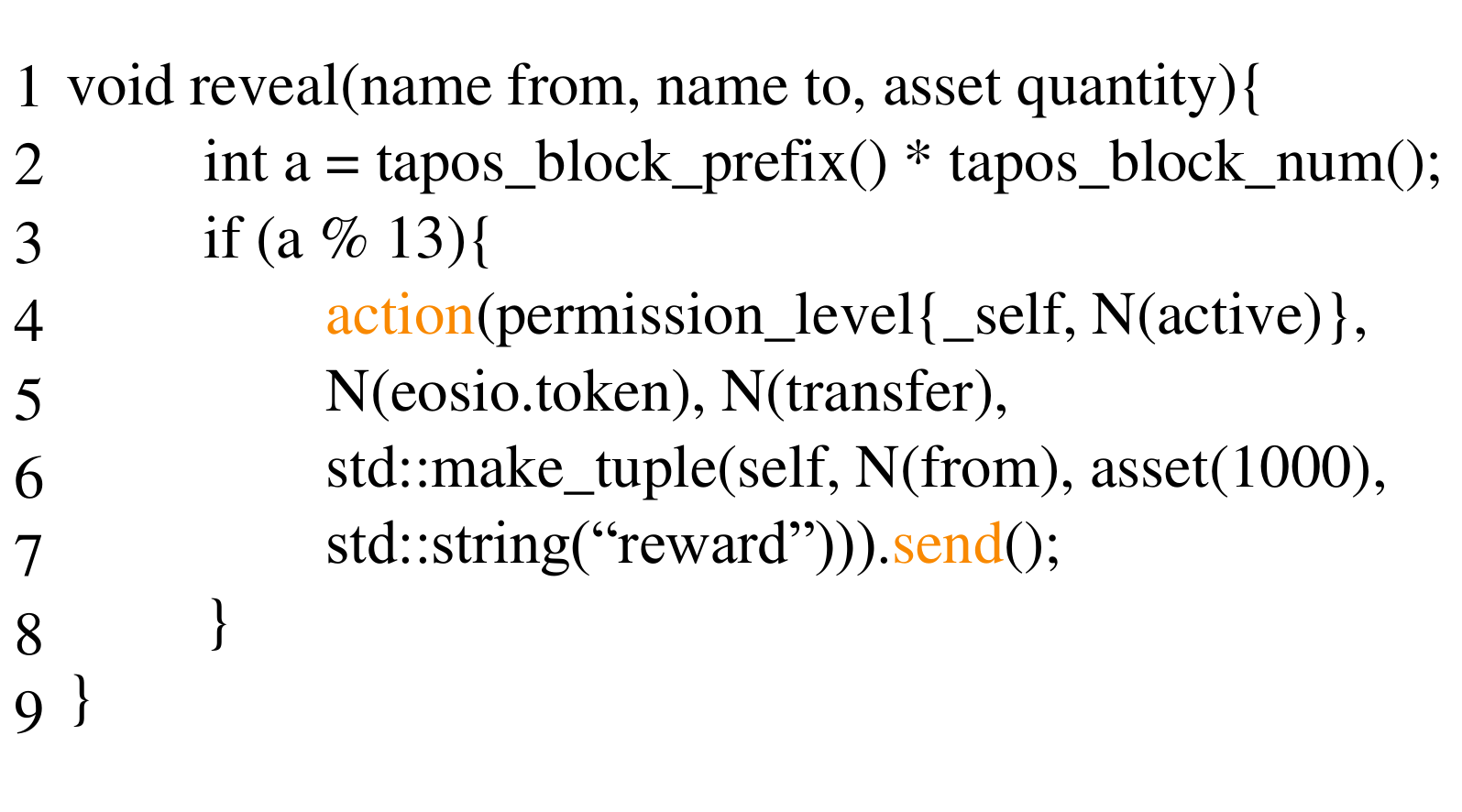} 
\caption{An example of Rollback vulnerabilities.}
\label{fig:rollexm}
\end{figure}
Fig. \ref{fig:rollexm} illustrates the $\textsl{reveal}$ function of a lottery smart contract. This function, which can be called by any user, generates a random number (Line 2) and initiates a transfer to the caller (Line 4-7) if the generated number meets specific conditions (Line 3). The transfer is implemented through $\textsl{action}(\ldots).\textsl{send()}$, which internally utilizes the $\textsl{send\_inline}$ library function. This implementation creates a vulnerability: an attacker can exploit it by deploying a malicious contract that executes two sequential inline actions - first calling the $\textsl{reveal}$ function, then immediately checking the contract's balance. Since these inline actions, along with the lottery contract's transfer action, are bundled into a single transaction, the attacker can strategically abort the transaction by deliberately failing the second action upon discovering an unsuccessful lottery attempt. Through this mechanism, the attacker can repeatedly attempt to win the lottery without risk, rolling back unsuccessful attempts until achieving a profitable outcome.

% An example of this vulnerability exists in some lottery contracts. Consider a contract where prize distribution occurs through inline actions for fund transfers. An attacker can craft an attack by implementing a contract that combines two inline actions within a single transaction: one to trigger the prize distribution function and another to monitor attacker's balance. By programming the second action to fail if no balance increase is detected, the attacker creates a mechanism where losing transactions are automatically rolled back, effectively guaranteeing they never lose.

\item \textit{Missing Permission Check.}
Contracts that inadequately implement authorization checks for sensitive operations create security gaps where attackers can bypass both contract-level user permissions and platform-level verification mechanisms. 

\item \textit{Integer Overflow.}
Similar to traditional binary code, since EOSIO platform smart contracts' WASM supports fixed-length integers, integer overflow vulnerabilities may occur if proper boundary are not implemented during integer calculations.
\end{itemize}

Note that this paper primarily focuses on vulnerabilities related to on-chain data, which current tools cannot accurately detect. 
Therefore, we exclude the detection of Fake EOS and Fake Receipt vulnerabilities, as these fall outside our scope and can be effectively identified using existing tools.

\section{On-chain Data API Emulation}

\subsection{The Necessity of Emulation}
\label{subsec:motivate}

Based on our comprehensive analysis of a real-world dataset comprising 5,602 WASM smart contracts, we conduct a systematic investigation of on-chain data API usage patterns. 
The results reveal that 3,579 contracts incorporate APIs for reading on-chain data, while 3,498 contracts utilize APIs for modifying on-chain data. 
Overall, 64.07\% (3589/5602) of the analyzed contracts demonstrate dependence on on-chain data APIs. 
% Through experimental analysis, we find that many EOSIO smart contracts, 64.07\% (3589/5602) in our real-world dataset, use on-chain data API functions. 
% This frequent interaction with on-chain data tables is necessitated due to the memory isolation between contract actions. Since each action initializes its own memory block for variable storage during execution, different actions cannot directly share variables through memory. Consequently, the popular method for transferring variables between actions is to persist them in on-chain data tables.

\begin{figure}
\centering
\includegraphics[scale=0.3]{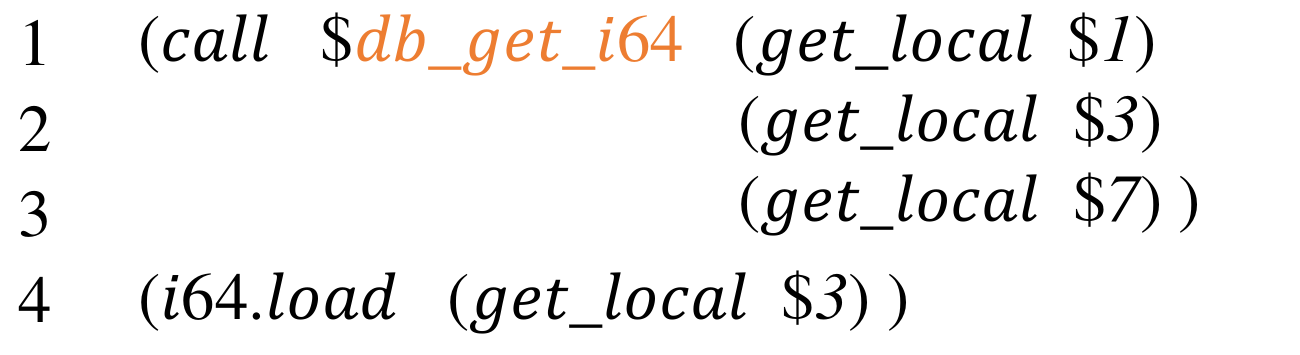} 
\caption{A fragment of a contract using on-chain data APIs.}
\label{fig:exm}
\end{figure}

The above findings suggest that the emulation of on-chain data APIs is relevant to the analysis of the majority of smart contracts.
However, we find that the existing coarse-grained emulation of those APIs may introduce significant inaccuracies for the analysis.
Taking the WASM contract fragment shown in the Fig. \ref{fig:exm} as an example, lines 1-3 demonstrate the process of retrieving data from the on-chain database. The code begins by obtaining three variables that serve as arguments for the $\textsl{db\_get\_i64}$ function: the on-chain data iterator $\textsl{get\_local \$1}$, the starting memory address for storing the retrieved data $\textsl{get\_local \$3}$, and the allocated memory length $\textsl{get\_local \$7}$. Upon execution, this function returns the actual length of the retrieved data. While existing methods perform symbolic execution of this code by representing the return value as either random or symbolic, they fail to simulate the crucial internal operation of storing chain data in memory. Consequently, when line 4 attempts to load data from the specified memory address, i.e., $\textsl{get\_local \$3}$, it encounters incorrect or null values. This limitation can compromise subsequent symbolic execution and impact vulnerability detection accuracy: for instance, overflow vulnerabilities may go undetected when data is incorrectly interpreted as null.

% 在如图所示的wasm合约片段中，第一行到第三行的代码是用于从链上数据库获取数据.
% 具体而言，该段代码首先栈中获取三个变量的取值，分为作为dbget函数的参数，表示链上数据的迭代器、链上数据取得后存放区域的内存首地址，存放区域的长度。
% 该函数执行完成后会返回数据的长度。
% 现有方法符号执行上述代码时，会以随机值或符号值表示返回值，但不模拟内部的操作，即将链上数据存于内存中。
% 因此，第4行代码从内存的对应地址加载数据时，只能获得一个错误的值或空值。
% 这将导致后续的代码符号执行结果都可能出现错误，最终可能影响到漏洞检测的结果。

% 为了解决上述问题，我们提出了细粒度模拟链上数据API的方法，将API的内存操作纳入考虑，从而提高链上数据相关漏洞的检测准确率。
% 然而，我们发现，细粒度模拟链上数据API可能增加分析的时间开销，甚至导致分析无法在规定时间内完成（见实验部分）。
% 因此，我们又进一步提出了以覆盖率导向的混合执行方法，通过在部分步骤引入具体执行提高执行效率，并通过多次循环保证代码覆盖率，最终实现分析准确率和效率的平衡。

To address the above issues, we propose a fine-grained emulation approach that precisely models on-chain data APIs by incorporating their memory operations, thereby enhancing the detection accuracy of vulnerabilities related to on-chain data.

\begin{table*}[]
\centering 
 \caption{The emulation of on-chain database API functions}
 \label{tab:emu}
\begin{tabular}{c|c|c}
\hline
API                    & Precondition   & Emulation Steps \\ \hline
\multirow{6}{*}{\begin{tabular}[c]{@{}c@{}}$\textsl{db\_find\_i64}(code,scope,$ \\ $table,id)$\end{tabular}}& \multirow{3}{*}{\begin{tabular}[c]{@{}c@{}}$ (\exists t \in s.tables.$ \\ $\exists i \in t.iters.\ get(i).key =id\wedge t.owner=code$ \\ $ \wedge t.scope = scope\wedge t.name = table)$ \end{tabular}} & \multirow{3}{*}{\begin{tabular}[c]{@{}c@{}} $ s.ret \leftarrow i $\end{tabular}}   \\   
                               &                                                                                     &   \\
                               &                                                                                     &   \\ \cline{2-3}
                               &\multirow{3}{*}{\begin{tabular}[c]{@{}c@{}}$ (\exists t \in s.tables.\ t.owner=$ \\ $code\wedge t.scope = scope\wedge t.name = table)$\\ $\wedge(\forall i \in t.iters.\ get(i).key \neq id) $ \end{tabular}} & \multirow{3}{*}{\begin{tabular}[c]{@{}c@{}} $s.ret \leftarrow t.iters[len(t.iters)-1] $\end{tabular}}   \\   
                               &                                                                                     &   \\
                               &                                                                                     &   \\  \hline
\multirow{4}{*}{\begin{tabular}[c]{@{}c@{}}$\textsl{db\_get\_i64}(iter,data,length)$\end{tabular}}& \multirow{2}{*}{\begin{tabular}[c]{@{}c@{}}$(type(iter) = \textsl{CON}) \wedge $ \\ $(\exists t \in s.tables.\ iter \in t.iters) \wedge (length\neq 0)$ \end{tabular}} & \multirow{2}{*}{\begin{tabular}[c]{@{}c@{}} $s.mem[data,data+length] \leftarrow$\\ $get(iter) $\end{tabular}}   \\   
                               &                                                                                     &   \\ \cline{2-3}
                               & \multirow{2}{*}{\begin{tabular}[c]{@{}c@{}}$(type(iter) = \textsl{CON}) \wedge $ \\ $(\exists t \in s.tables.\ iter \in t.iters) \wedge (length = 0)$ \end{tabular}} & \multirow{2}{*}{\begin{tabular}[c]{@{}c@{}} $s.ret \leftarrow len(get(iter)) $\end{tabular}}   \\   
                               &                                                                                     &   \\ \hline
                               % \\ \cline{2-3} 
                               % & \multirow{2}{*}{\begin{tabular}[c]{@{}c@{}}$type(iter) = \textsl{CON}$ \\ $\forall t \in s.tables.\ iter \notin t.iters $ \end{tabular}} & \multirow{2}{*}{\begin{tabular}[c]{@{}c@{}}revert \end{tabular}}     \\   
                               % &                                                                                     &    
\multirow{6}{*}{\begin{tabular}[c]{@{}c@{}}$\textsl{db\_next\_i64}(iter,prim)$\end{tabular}}& \multirow{3}{*}{\begin{tabular}[c]{@{}c@{}}$(type(iter) = \textsl{CON}) \wedge$ \\ $(\exists t \in s.tables.\ iter \in t.iters \wedge$ \\ $t.iters.index(iter) \neq len(t.iters) - 1)$ \end{tabular}} & \multirow{3}{*}{\begin{tabular}[c]{@{}c@{}} $i \leftarrow t.iters.index(iter) + 1$ \\ $\curvearrowright\  prim \leftarrow get(i).key $\end{tabular}}   \\   
                               &                                                                                     &   \\
                               &                                                                                     &   \\ \cline{2-3}
                               & \multirow{3}{*}{\begin{tabular}[c]{@{}c@{}}$(type(iter) = \textsl{CON}) \wedge$ \\ $(\exists t \in s.tables.\ iter \in t.iters \wedge$ \\ $t.iters.index(iter) = len(t.iters) - 1)$ \end{tabular}} & \multirow{3}{*}{\begin{tabular}[c]{@{}c@{}} $ \emptyset $\end{tabular}}   \\   
                               &                                                                                     &   \\
                               &                                                                                     &   \\ \hline
                               %  \\ \cline{2-3} 
                               % & \multirow{2}{*}{\begin{tabular}[c]{@{}c@{}}$type(iter) = \textsl{CON}$ \\ $\forall t \in s.tables.\ iter \notin t.iters $ \end{tabular}} & \multirow{2}{*}{\begin{tabular}[c]{@{}c@{}}revert \end{tabular}}     \\   
                               % &                                                                                     &  
\multirow{7}{*}{\begin{tabular}[c]{@{}c@{}}$\textsl{db\_lowerbound\_i64}(code,$ \\ $scope,table,id)$\end{tabular}}& \multirow{4}{*}{\begin{tabular}[c]{@{}c@{}}$ (\exists t \in s.tables.$ \\ $\exists its \subset t.iters.\ \forall i \in its.\ get(i).key \geq id$ \\ $\wedge t.owner=code \wedge t.scope = scope$ \\ $ \wedge t.name = table)$ \end{tabular}} & \multirow{4}{*}{\begin{tabular}[c]{@{}c@{}} $ s.ret \leftarrow min(its) $\end{tabular}}   \\   
                               &                                                                                     &   \\
                               &                                                                                     &   \\
                               &                                                                                     &   \\ \cline{2-3}
                               &\multirow{3}{*}{\begin{tabular}[c]{@{}c@{}}$ (\exists t \in s.tables.\ t.owner=$ \\ $code\wedge t.scope = scope\wedge t.name = table)$\\ $\wedge(\forall i \in t.iters.\ get(i).key < id) $ \end{tabular}} & \multirow{3}{*}{\begin{tabular}[c]{@{}c@{}} $s.ret \leftarrow t.iters[len(t.iters)-1] $\end{tabular}}   \\   
                               &                                                                                     &   \\
                               &                                                                                     &   \\  \hline
\multirow{7}{*}{\begin{tabular}[c]{@{}c@{}}$\textsl{db\_upperbound\_i64}(code,$ \\ $scope,table,id)$\end{tabular}}& \multirow{4}{*}{\begin{tabular}[c]{@{}c@{}}$ (\exists t \in s.tables.$ \\ $\exists its \subset t.iters.\ \forall i \in its.\ get(i).key > id$ \\ $\wedge t.owner=code \wedge t.scope = scope$ \\ $ \wedge t.name = table)$ \end{tabular}} & \multirow{4}{*}{\begin{tabular}[c]{@{}c@{}} $ s.ret \leftarrow min(its) $\end{tabular}}   \\   
                               &                                                                                     &   \\
                               &                                                                                     &   \\
                               &                                                                                     &   \\ \cline{2-3}
                               &\multirow{3}{*}{\begin{tabular}[c]{@{}c@{}}$ (\exists t \in s.tables.\ t.owner=$ \\ $code\wedge t.scope = scope\wedge t.name = table)$\\ $\wedge(\forall i \in t.iters.\ get(i).key \leq id) $ \end{tabular}} & \multirow{3}{*}{\begin{tabular}[c]{@{}c@{}} $s.ret \leftarrow t.iters[len(t.iters)-1] $\end{tabular}}   \\   
                               &                                                                                     &   \\
                               &                                                                                     &   \\   \hline 
\multirow{6}{*}{\begin{tabular}[c]{@{}c@{}}$\textsl{db\_previous\_i64}(iter,prim)$\end{tabular}}& \multirow{3}{*}{\begin{tabular}[c]{@{}c@{}}$(type(iter) = \textsl{CON}) \wedge$ \\ $(\exists t \in s.tables.\ iter \in t.iters \wedge$ \\ $t.iters.index(iter) \neq 0)$ \end{tabular}} & \multirow{3}{*}{\begin{tabular}[c]{@{}c@{}} $i \leftarrow t.iters.index(iter) - 1$ \\ $\curvearrowright\  prim \leftarrow get(i).key $\end{tabular}}   \\   
                               &                                                                                     &   \\
                               &                                                                                     &   \\ \cline{2-3}
                               & \multirow{3}{*}{\begin{tabular}[c]{@{}c@{}}$(type(iter) = \textsl{CON}) \wedge$ \\ $(\exists t \in s.tables.\ iter \in t.iters \wedge$ \\ $t.iters.index(iter) = 0)$ \end{tabular}} & \multirow{3}{*}{\begin{tabular}[c]{@{}c@{}} $ \emptyset $\end{tabular}}   \\   
                               &                                                                                     &   \\
                               &                                                                                     &   \\ \hline

\end{tabular}
\end{table*}

\subsection{Emulation of On-chain data API functions}
\label{subsec:basicemu}

In this section, we present our emulation of the on-chain data APIs.
First, we introduce the notation used in our symbolic execution engine. 
A table in an on-chain database can be formally denoted as $<owner,scope,name,iters,datas>$, where $owner$, $scope$ and $name$ denote the information to uniquely identify this table, $iters$ represents the set of iterators associated with the table, and $datas$ represents the set of key-value pairs $<key,value>$ stored within the table, as introduced in Section \ref{sec:back}. The on-chain database itself is composed of multiple tables and can be represented as $tabs$, which is a set of tables. 
The complete execution state is denoted by $s=<tabs,mem,\theta,ret>$, where $mem$ represents the memory during execution, $\theta$ denotes the set of constraints corresponding to the current execution branch, and $ret$ indicates the return value of the currently executed function. 
To handle multiple execution paths, the engine maintains a set $S$ of states representing different execution branches.
In addition, we implemente several functions: The $get$ function accepts an iterator as input and retrieves the corresponding row of data. The $index$ function takes both an ordered set and an element as inputs, returning the element's position within that set. The $len$ function calculates the total number of elements in an ordered set, while the $add$ and $remove$ functions handle the insertion and deletion of elements from the set respectively.
The $min$ function outputs the smallest element in a set.

% The execution state is denoted by s=<cdatas,memory,constraints,returns>, where cdatas represents the data stored in the current chain, memory represents the system memory during execution, constraints denotes the set of conditions corresponding to the current execution branch, and returns indicates the return value of the currently executed function. The engine maintains a set S of states to represent different execution branches.

% % 根据第2章中链上数据结构的介绍，我们用iters,datas表示一个表格，其中iters表示表格对应的迭代器的集合，datas表示表格中的键值对<key,value>的集合。
% 而链上数据库则用表格的集合tables表示。
% 符号执行的状态s=<tables,mem,\theta,ret>

% According to the introduction of on-chain table in Section 2, we denote a table by <iters,datas>, where iters denotes the set of iterators corresponding to the table and datas denotes the set of key-value pairs <key,value> in the table.
% And a on-chain database is represented by tabs, a set of tables.
% The execution state is denoted by s=<tabs,mem,theta,ret>, where mem represents the system memory during execution, theta denotes the set of conditions corresponding to the current execution branch, and ret indicates the return value of the currently executed function. The engine maintains a set S of states to represent different execution branches.

% cdatas表示为tables,keys,iters，其中三个元素分别表示链上的数据表名、指向不同数据行的索引以及迭代器。

Based on this notation, we emulate the on-chain data API functions~\cite{apis} shown in Table \ref{tab:emu}.
In the table, Column 2 specifies the preconditions required for function execution, and Column 3 demonstrates our emulation steps for these conditions, with $\curvearrowright$ denoting the sequential dependencies between steps.
Here we take the APIs $\textsl{db\_*\_i64}$ for 64-bit integer index tables as examples, and the emulation of other APIs follows similar steps.
We also omit the emulation of the rollback caused by the exception of execution of APIs for brevity.

% 注意，我们这里展示的是与64位索引表格相关的部分API，对于其他类型表格的模拟与该类API的模拟类似，因此不再赘述。
% 同时由于篇幅原因，我们省略了API内部执行出错回滚的情况模拟。

% Here we take the operations db\_*\_i64 as examples, and the emulation of operations of other types follow similar patterns:

% 64-bit integer index table.

\textbf{db\_find\_i64(code, scope, table, id).}
This function locates the iterator for a specific data row using four parameters:
\begin{enumerate}
\item $code$: the table owner's name.
\item $scope$: the table's scope definition.
\item $table$: the table name identifier.
\item $id$: the target row's index.
\end{enumerate}

The function's behavior varies based on three scenarios: if the specified table does not exist on the blockchain, the function fails and triggers a transaction rollback. If the table exists but lacks a row matching the provided $id$, it returns the table's end-row iterator. If both the table and the row with the matching $id$ exist, it returns that row's iterator.

Thus, we implement distinct emulation strategies based on the data existence:

For $id$ with existing corresponding rows, the matching row's iterator is assigned to the $ret$ of the current execution state.
For $id$ without matching rows, the table's last row iterator is assigned to $ret$.
% For symbolic $id$ values, the engine treats $id$ as the table index, adds corresponding constraints to $\theta$, and generates new execution states in $S$ to represent different execution branches.

\textbf{db\_get\_i64(iter, data, length).}
This function retrieves and stores row data in memory based on three parameters:
\begin{enumerate}
\item $iter$: iterator of the target row.
\item $data$: memory buffer location for data storage.
\item $length$: buffer size.
\end{enumerate}
The function's operation depends on the $length$ parameter: when non-zero, it retrieves and stores the row data in the specified memory buffer ($data$ to $data+length$); when zero, it only returns the data length without performing storage operations. 

We emulate both scenarios: storing fetched data into $s.mem[data,data+length]$ for non-zero $length$, or assigning $len(get[i])$ to $ret$  otherwise. 
% For symbolic iterators, the engine assumes that $iter$ equals to each chain table iterators, adds constraints to $\theta$, processes data storage, and generates various execution states for $S$.

\textbf{db\_next\_i64(iter, prim).}
This function processes the row data corresponding to iterator $iter$ and stores the next row's index in $prim$. 
If $iter$ is not pointing to the last row of a table, the function performs the retrieving and storing operations.
Otherwise, it performs no operations, leaving $prim$ unmodified. 
We emulate both scenarios: storing the fetched iterator into $prim$ for $iter$ that is not the last one, or doing nothing otherwise. 
% and handles symbolic iterators similarly to the $\textsl{db\_get\_i64}$ function, treating $iter$ as corresponding to individual chain table iterators for separate processing.

Additionally, $\textsl{db\_lowerbound\_i64}(code,scope,table,id)$ and $\textsl{db\_upperbound\_i64}(code,scope,table,id)$ functions retrieve table rows matching lowerbound or upperbound conditions, i.e., the first table row with the lowest primary key that is $\geq id$ or $>id$.
These functions follow simulation patterns similar to $\textsl{db\_find\_i64}$. 
The $\textsl{db\_previous\_i64}$ function, which retrieves the previous row's iterator, follows emulation steps similar to $\textsl{db\_next\_i64}$.

\section{\ourtool}

\begin{table*}[]
\centering 
 \caption{The emulation of on-chain database API functions under Concolic Execution}
 \label{tab:concolicemu}
\begin{tabular}{c|c|c}
\hline
API                    & Precondition   & Emulation Steps \\ \hline
\multirow{10}{*}{\begin{tabular}[c]{@{}c@{}}$\textsl{db\_find\_i64}(code,scope,$ \\ $table,id)$\end{tabular}}& \multirow{3}{*}{\begin{tabular}[c]{@{}c@{}}$(type(id) = \textsl{CON}) \wedge (\exists t \in s.tables.$ \\ $\exists i \in t.iters.\ get(i).key =id\wedge t.owner=$ \\ $code \wedge t.scope = scope\wedge t.name = table)$ \end{tabular}} & \multirow{3}{*}{\begin{tabular}[c]{@{}c@{}} $ s.ret \leftarrow i $\end{tabular}}   \\   
                               &                                                                                     &   \\
                               &                                                                                     &   \\ \cline{2-3}
                               &\multirow{3}{*}{\begin{tabular}[c]{@{}c@{}}$(type(id) = \textsl{CON}) \wedge (\exists t \in s.tables.$ \\ $t.owner=code\wedge t.scope = scope\wedge t.name$\\ $ = table)\wedge(\forall i \in t.iters.\ get(i).key \neq id) $ \end{tabular}} & \multirow{3}{*}{\begin{tabular}[c]{@{}c@{}} $s.ret \leftarrow t.iters[len(t.iters)-1] $\end{tabular}}   \\   
                               &                                                                                     &   \\
                               &                                                                                     &   \\  \cline{2-3}
                               &\multirow{4}{*}{\begin{tabular}[c]{@{}c@{}}$(type(id) = \textsl{SYM})\wedge (\exists t \in s.tables.$ \\ $ t.owner=code\wedge t.scope = scope\wedge$ \\ $ t.name = table)$ \end{tabular}} & \multirow{4}{*}{\begin{tabular}[c]{@{}c@{}}$S.remove(s)\ \curvearrowright\  \forall i \in t.iters.$\\ $\{s' \leftarrow s \ \curvearrowright\   s'.\theta.add($ \\ $get(i).key=id) \ \curvearrowright\ s'.ret \leftarrow i \ $ \\ $\curvearrowright  S.add(s')\}$\end{tabular}} \\
                               &                                                                                     &   \\ 
                               &                                                                                     &   \\ 
                               &                                                                                     &   \\ \hline
\multirow{13}{*}{\begin{tabular}[c]{@{}c@{}}$\textsl{db\_get\_i64}(iter,data,length)$\end{tabular}}& \multirow{2}{*}{\begin{tabular}[c]{@{}c@{}}$(type(iter) = \textsl{CON}) \wedge (\exists t \in s.tables.$ \\ $ iter \in t.iters) \wedge (length\neq 0)$ \end{tabular}} & \multirow{2}{*}{\begin{tabular}[c]{@{}c@{}} $s.mem[data,data+length] \leftarrow$ \\ $ get(iter) $\end{tabular}}   \\   
                               &                                                                                     &   \\ \cline{2-3}
                               & \multirow{5}{*}{\begin{tabular}[c]{@{}c@{}} $(type(iter) = \textsl{SYM}) \wedge (length\neq 0)$\end{tabular}} & \multirow{5}{*}{\begin{tabular}[c]{@{}c@{}}$S.remove(s)\ \curvearrowright \ \forall t \in s.tables.$ \\ $\forall i \in t.iters.\{s' \leftarrow s \ \curvearrowright\  s'.\theta.add($ \\ $iter=i) \ \curvearrowright\ s'.mem[data,data+$ \\ $length] \leftarrow get(i) \ \curvearrowright\  S.add(s')\}$\end{tabular}}   \\   
                               &                                                                                     &   \\ 
                               &                                                                                     &   \\ 
                               &                                                                                     &   \\ 
                               &                                                                                     &   \\ \cline{2-3}
                               & \multirow{2}{*}{\begin{tabular}[c]{@{}c@{}}$(type(iter) = \textsl{CON}) \wedge (\exists t \in s.tables.$ \\ $ iter \in t.iters) \wedge (length = 0)$ \end{tabular}} & \multirow{2}{*}{\begin{tabular}[c]{@{}c@{}} $s.ret \leftarrow len(get(iter)) $\end{tabular}}   \\   
                               &                                                                                     &   \\ \cline{2-3}
                               & \multirow{4}{*}{\begin{tabular}[c]{@{}c@{}} $(type(iter) = \textsl{SYM}) \wedge (length= 0)$\end{tabular}} & \multirow{4}{*}{\begin{tabular}[c]{@{}c@{}}$S.remove(s)\ \curvearrowright \ \forall t \in s.tables.$ \\ $ \forall i \in t.iters.\ \{s' \leftarrow s \ \curvearrowright\  s'.\theta.add($ \\ $iter=i) \ \curvearrowright\  s'.ret \leftarrow len(get(i))$ \\ $ \curvearrowright\  S.add(s')\}$\end{tabular}}   \\   
                               &                                                                                     &   \\ 
                               &                                                                                     &   \\ 
                               &                                                                                     &   \\\hline
                               % \\ \cline{2-3} 
                               % & \multirow{2}{*}{\begin{tabular}[c]{@{}c@{}}$type(iter) = \textsl{CON}$ \\ $\forall t \in s.tables.\ iter \notin t.iters $ \end{tabular}} & \multirow{2}{*}{\begin{tabular}[c]{@{}c@{}}revert \end{tabular}}     \\   
                               % &                                                                                     &    
\multirow{11}{*}{\begin{tabular}[c]{@{}c@{}}$\textsl{db\_next\_i64}(iter,prim)$\end{tabular}}& \multirow{3}{*}{\begin{tabular}[c]{@{}c@{}}$(type(iter) = \textsl{CON}) \wedge$ \\ $(\exists t \in s.tables.\ iter \in t.iters \wedge$ \\ $t.iters.index(iter) \neq len(t.iters) - 1)$ \end{tabular}} & \multirow{3}{*}{\begin{tabular}[c]{@{}c@{}} $i \leftarrow t.iters.index(iter) + 1 \ \curvearrowright$\\ $  prim \leftarrow get(i).key $\end{tabular}}   \\   
                               &                                                                                     &   \\
                               &                                                                                     &   \\ \cline{2-3}
                               & \multirow{3}{*}{\begin{tabular}[c]{@{}c@{}}$(type(iter) = \textsl{CON}) \wedge$ \\ $(\exists t \in s.tables.\ iter \in t.iters \wedge$ \\ $t.iters.index(iter) = len(t.iters) - 1)$ \end{tabular}} & \multirow{3}{*}{\begin{tabular}[c]{@{}c@{}} $ \emptyset $\end{tabular}}   \\   
                               &                                                                                     &   \\
                               &                                                                                     &   \\ \cline{2-3}
                               & \multirow{5}{*}{\begin{tabular}[c]{@{}c@{}} $type(iter) = \textsl{SYM}$\end{tabular}} & \multirow{5}{*}{\begin{tabular}[c]{@{}c@{}}$S.remove(s)\ \curvearrowright\  \forall t \in s.tables.$ \\ $\forall i \in t.iters[0,len(t.iters)-2]. $ \\ $\{ s' \leftarrow s \ \curvearrowright\ i' \leftarrow t.iters.index(iter) $ \\ $+ 1  \curvearrowright\  prim \leftarrow get(i').key \ \curvearrowright$ \\ $  s'.\theta.add(iter=i) \ \curvearrowright   S.add(s')\}$\end{tabular}}   \\   
                               &                                                                                     &  \\
                               &                                                                                     &  \\
                               &                                                                                     &  \\
                               &                                                                                     &  \\ \hline

\end{tabular}
\end{table*}

\subsection{Efficiency Challenge}

While fine-grained emulation of on-chain APIs enhances analytical accuracy, our investigation reveals that incorporating such emulation mechanisms significantly impacts the computational performance of analyzers. 
We implement the aforementioned on-chain API emulation by extending the existing symbolic execution tool WANA, and conduct extensive performance evaluation experiments across 5,602 real-world smart contracts.
The experimental results demonstrate that the modified tool requires an average of 1,482 seconds to complete the analysis of a single contract. 
Within a 1,800-second timeout threshold, the tool fails to complete analysis for 16.21\% (908/5,602) of the contracts. 
Furthermore, even when extending the timeout threshold to 3,600 and 7,200 seconds, the tool still exhibits analysis incompletion rates of 12.71\% (712/5,602) and 1.98\% (111/5,602), respectively.
These results demonstrate that simply incorporating on-chain API emulation onto existing tools proves impractical for the analysis of numerous real-world contracts, highlighting the need for a more sophisticated approach to balance efficiency and accuracy.

% r investigation reveals that incorporating on-chain API simulation significantly impacts the efficiency of analysis tools. 
% To quantify this impact, we implement on-chain API emulation based on WANA and conduct comparative evaluations on 5,602 real-world contracts. 
% % While WANA successfully completes verification for all contracts within a 3,600-second time limit per contract, the modified tool encounters timeout issues on xxx contracts. 
% These results demonstrate that simply retrofitting on-chain API emulation onto existing tools proves impractical for real-world contract analysis, highlighting the need for a more sophisticated approach to balance efficiency and accuracy.

To address the above challenge, we employ concolic execution, a technique widely used for analyzing various types of programs~\cite{vidal2014concolic}\cite{liang2020practical}\cite{ai2020novel}. Unlike purely symbolic execution, concolic execution's key advantage lies in its use of concrete values (both data and address), enabling precise reasoning about complex data structures while simplifying constraints~\cite{majumdar2007hybrid}. Building on this foundation, we propose WACANA, whose architecture is illustrated in Fig. \ref{fig:design}. WACANA implements fine-grained emulation of on-chain data APIs and leverages multi-round concolic execution for achieving an optimal balance between detection accuracy and efficiency. The following sections detail the core components of our approach.

\subsection{Multi-Round Concolic Execution}

% While simulating the internal operations of the on-chain API enhances analysis accuracy, it increases both the number of constraints and symbolic variables that require resolution. This complexity often leads to additional symbolic execution branches, resulting in substantial computational overhead. To optimize performance without compromising analysis effectiveness, we implement a strategy of concretizing on-chain data, which reduces the number of symbolic variables and thereby improves the overall efficiency of our analysis process.

Our concolic execution method combines concrete and symbolic execution by delegating the generation and modification of on-chain data to an actual virtual machine while symbolically executing contract actions. 
This hybrid approach strategically concretizes frequently accessed on-chain data, significantly reducing both the symbolic variables during constraint solving and the number of branches requiring exploration.

Our tool operates through a hybrid execution loop, beginning with the initialization of an empty on-chain database. The execution process, illustrated in Fig. \ref{fig:design}, consists of two main phases: First, the contract functions undergo separate symbolic execution to generate a collection of result states. Subsequently, one result state is selected to generate an actual transaction, which is then executed on the test chain. This execution updates the on-chain database, providing new data for the next execution round.

\textbf{State Selection.} The selection of result states is driven by instruction coverage growth. During execution, our symbolic execution engine records both the sequence of executed instructions and the set of required constraints, linking them to the result states. For each state, we calculate its incremental coverage by comparing the number of newly covered instructions with the previous state. The execution terminates if no state shows increased coverage. Otherwise, we select the state with the highest coverage gain for further processing. In cases where multiple states achieve equal coverage increases, we prioritize states containing instructions updating on-chain data, as these facilitate the generation of new on-chain data. Our experimental comparison between this coverage-guided approach and random selection in Section \ref{subsec:componenttest} confirms that the coverage-guided strategy achieves superior instruction coverage and analysis accuracy.

\textbf{Transaction Generation.} Once a state is selected, we identify the target function and necessary triggering action. This process involves determining the function name, parameter types, and values. Since WASM contracts are not human-readable, we employ a tool to generate an Application Binary Interface (ABI) file~\cite{apis3}, which presents the contract's functions and parameter types in JSON format. We then analyze the instruction sequences associated with the selected state to identify the target function and derive parameter data types from the ABI file. Parameter values are generated by solving the state's constraint set. For simple data types (e.g., fixed-length integers or floating-point numbers), constraint solving directly yields parameter values. Complex data types (e.g., numeric assets or structures) require a more sophisticated three-step approach:
\begin{enumerate}
\item Examine consecutive memory values at the symbolic address location based on the parameter's data type, using these values when specifically defined
\item Identify and solve constraints for individual parameter elements, then combine the resulting values
\item Generate random element values when constraints are unavailable
\end{enumerate}
The final parameter values are packaged into a transaction for test chain execution.

\textbf{Transaction Execution.} Our test chain manages all transaction executions, which is implemented based on the WASAI tool. The chain's Virtual Machines generate log files documenting on-chain data changes resulting from transaction execution. The execution loop terminates when a generated transaction fails to produce any changes in the on-chain data.

\subsection{On-Chain Data API Emulation under Concolic Execution}

Based on the implementation of multiple rounds of concolic execution, we need to redesign the emulation method of the on-chain data APIs to accommodate the alternation of concrete and symbolic execution.
According to our basic emulation methods mentioned in Section \ref{subsec:basicemu}, we develop an approach for concolic execution scenarios, partly shown in TABLE \ref{tab:concolicemu}.
We define the $type$ function, which outputs $\textsl{SYM}$ or $\textsl{CON}$ to indicate whether the given value is symbolic or concrete, respectively.
The key difference lies in our parameter handling: during the execution, the emulation of the functions with concrete parameters follow the basic emulation steps, while the functions with symbolic parameters undergo comprehensive case discussion.
For instance, when handling a $\textsl{db\_get\_i64}$ function with a symbolic $iterator$, 
We assume that $iterator$ is equal to each iterator in the corresponding table respectively and generate different states $s'$ based on the different assumptions and add them into the set $S$.
In this way, we form multiple branches of execution and thus discuss possible outcomes for different values of $iterator$.
Similarly, for $\textsl{db\_find\_i64}$ function with symbolic $id$, we consider each possible table index, incorporating appropriate constraints and generating corresponding execution states. 
This approach extends to all API emulations, ensuring thorough coverage of possible execution paths.

\subsection{Vulnerabilities Detector}
\label{subsec:vuldetect}

Similar to existing methods \cite{he2021eosafe}\cite{jiang2021wana}, \ourtool uses a detector to identify vulnerability patterns in paths during symbolic execution. Any path matching these patterns is flagged as a potential vulnerability.
The detection patterns in our tool for different types of vulnerabilities, based on expert knowledge, are categorized as follows:

\textbf{Blockchain-info Dependency.}
A Blockchain-info Dependency vulnerability is detected when two conditions are met in an execution path: 
\begin{enumerate}
\item The presence of calls to blockchain information-related library functions~\cite{apis2}, e.g., $\textsl{tapos\_block\_num}$.
\item The use of this blockchain information as a conditional parameter that determines the execution of sensitive operations. 
\end{enumerate}
Such a dependency creates potential security risks as the blockchain information could be manipulated or predicted by malicious actors to influence the execution of sensitive operations.

\textbf{Rollback.}
A rollback vulnerability is detected when a path satisfies two conditions:
\begin{enumerate}
\item The use of blockchain information (such as timestamp or block number) as an operand in remainder instructions ($\textsl{rem}$).
\item The presence of calls to inline action library functions.
\end{enumerate}
This vulnerability typically appears in gambling smart contracts, where the blockchain state is commonly used as an operand in remainder instructions. 
So we propose the first condition.
What's more, The second condition indicates that the path's execution completion depends on other contracts, potentially allowing attackers to exploit rollback mechanisms.

\textbf{Missing Permission Check.}
A contract is considered vulnerable when a path performs sensitive operations without calling the appropriate permission-checking library functions. Sensitive operations include updates and deletions to the on-chain database, as well as calls to other contracts using official library functions. These operations should be restricted to authorized users or administrators rather than being publicly accessible.

\textbf{Integer Overflow.}
When a path executes fixed-length integer operations (add, subtract, or multiply), and the result is a symbolic value, \ourtool adds a constraint $r = bound$ to the path constraints set, where $r$ represents the instruction result and $bound$ represents the fixed-length integer boundary. If this constraint can be satisfied, an integer overflow vulnerability exists.

\section{Experiment}

\subsection{Setup}

\textbf{Datasets.} We construct a vulnerability dataset building upon WANA's original collection of 84 smart contracts~\cite{wanadata}. To expand this dataset, we systematically modify the contracts by either implementing vulnerability fixes or introducing specific vulnerabilities through controlled injections.
The vulnerability modification process follows these specific approaches for each type:

\begin{itemize}

\item \textit{Blockchain-info Dependency.} Removal of these vulnerabilities involved eliminating sensitive operations that are conditional on block-related information. To create vulnerable variants, we modify the seeds in the functions generating random numbers to depend on block timestamps or block header variables.

\item \textit{Rollback.} We address rollback vulnerabilities by replacing money transfer operations that used send\_inline and send\_context\_free\_inline library functions with send\_defered implementations. For vulnerability injection, we incorporate functions similar to those illustrated in Fig. \ref{fig:rollexm} into random contracts.

\item \textit{Missing Permission Check.} For vulnerability mitigation, we identify instructions associated with sensitive operations (as mentioned in section \ref{subsec:vuldetect}) and insert require\_auth calls before each operation. Conversely, to create vulnerable variants, we randomly remove existing require\_auth calls from certain contracts.

\item \textit{Integer Overflow.} To eliminate integer overflow vulnerabilities, we implement overflow-checking statements for all arithmetic operations (additions, subtractions, and multiplications) involving fixed-length parameters. To generate vulnerable versions, we identify functions with fixed-length integer parameters and strategically insert operations that could trigger overflows.

\end{itemize}

Following the dataset extension, we conduct manual verification of all modifications. The resulting \textit{vulnerability dataset} comprises 133 smart contracts with various vulnerability types: 57 contracts with missing permission check vulnerabilities, 35 with rollback vulnerabilities, 7 with blockchain-info dependency vulnerabilities, and 53 with integer overflow vulnerabilities. Note that individual contracts may contain multiple types of vulnerabilities.

Furthermore, we construct a dataset from the XBlock-EOS dataset~\cite{zheng2021xblock}, containing 55,735 smart contracts deployed on 5,714 different addresses in the EOS chain. 
We select the latest version of the smart contract corresponding to each address and then remove 112 empty contracts, ending up with a \textit{real-world dataset} of 5,602 contracts.

The source code of \ourtool please refer to \cite{ourcode}, and the above datasets are also open-sourced at \cite{ourdata}. 

\textbf{Environment.} We experiment on a server with Intel Xeon Gold 5215 2.50GHz CPU, 128G memory and Ubuntu 18.04 (64-bit).

\textbf{Questions.} Based on the above datasets and environment, we perform experimental evaluations, aiming to answer three key research questions:

\textit{RQ1.} How does our tool's vulnerability detection capability compare to existing representative analyzers for WASM contracts?

\textit{RQ2.} Can our tool effectively detect vulnerabilities in real-world WASM contracts?

\textit{RQ3.} Do our tool's core components, including on-chain data API emulation, concolic execution, and the coverage-guided loop, contribute to the vulnerability detection process in smart contracts?

% Please add the following required packages to your document preamble:
% \usepackage{multirow}
\begin{table}[]
\centering 
 \caption{A comparison of representative analyzers for WASM smart contracts on \textit{vulnerability dataset}. }
 \label{tab:comp}
\begin{tabular}{c|c|c|c|c}
\hline
Vulnerability                             & Tool      & Precision & Recall   & F1-score \\ \hline
\multirow{3}{*}{\begin{tabular}[c]{@{}c@{}}\textit{Missing Permission}\\ \textit{Check}\end{tabular}} & WASAI     & 75.00\%   & 15.79\%  & 0.26     \\ \cline{2-5} 
                                          & WANA      & /         & /        & /        \\ \cline{2-5} 
                                          & \ourtool & 100.00\%  & 92.98\%  & 0.96     \\ \hline
\multirow{3}{*}{\textit{Rollback}}                 & WASAI     & 100.00\%  & 20.00\%  & 0.33     \\ \cline{2-5} 
                                          & WANA      & /         & /        & /        \\ \cline{2-5} 
                                          & \ourtool & 100.00\%  & 85.71\%  & 0.92     \\ \hline
\multirow{3}{*}{\begin{tabular}[c]{@{}c@{}}\textit{Blockchain-info}\\ \textit{Dependency}\end{tabular}}      & WASAI     & 60.00\%   & 42.86\%  & 0.50     \\ \cline{2-5} 
                                          & WANA      & 100.00\%  & 57.14\%  & 0.72     \\ \cline{2-5} 
                                          & \ourtool & 100.00\%  & 100.00\% & 1.00     \\ \hline
\multirow{3}{*}{\textit{Integer Overflow}}                 & WASAI     & /         & /        & /        \\ \cline{2-5} 
                                          & WANA      & /         & /        & /        \\ \cline{2-5} 
                                          & \ourtool & 100.00\%  & 77.36\%  & 0.87     \\ \hline
\end{tabular}
\end{table}

\begin{table}[]
\centering 
 \caption{A comparison of the efficiency of representative analyzers for WASM smart contracts on \textit{vulnerability dataset}. }
 \label{tab:timecomp}
\begin{tabular}{c|c|c|c|c}
\hline
Tool      & Avg time & Min time & Max time & Medium time \\ \hline
WANA      & 9.26s    & 0.13s    & 314.91s  & 0.30s       \\ \hline
WASAI     & 54.67s   & 0.19s    & 301.92s  & 28.52s      \\ \hline
\ourtool & 498.03s  & 1.52s    & 3579.00s & 33.65s      \\ \hline
\end{tabular}
\end{table}

\subsection{The experiments comparing \ourtool and existing representative tools}

To answer RQ1, we evaluate our tool against two representative and open-source tools, which use different approaches: WANA~\cite{wanacode}, which employs symbolic execution, and WASAI~\cite{wasaicode}, which utilizes fuzzy testing techniques. Our evaluation use standard metrics including precision ($P=\frac{TP}{TP+FP}$), recall ($R=\frac{TP}{TP+FN}$), and F1 score($F1=\frac{2*P*R}{P+R}$), where F1 represents a balanced measure between recall and precision.
Here, $TP$ means that a vulnerable contract is correctly identified, $TN$ means that a contract without vulnerability is identified as secure, $FP$ means that a secure contract is incorrectly identified as vulnerable, and $FN$ means that a vulnerable contract is not detected.

The detailed vulnerability detection results are presented in TABLE \ref{tab:comp}. 
Our tool demonstrates superior performance across all metrics, achieving 100\% accuracy on the vulnerability dataset with higher recall and F1 scores across different vulnerability types. 
This enhanced performance can be attributed to our tool's more precise modeling of the on-chain database mechanism, which effectively reduces both false negatives and false positives in the detection of vulnerabilities related to on-chain data.
While our tool's recall rate falls short of 100\%, manual analysis of false negatives revealed that these instances stem from our concolic execution-based approach, which may not cover all path branches. 
Nevertheless, our method produces fewer false negatives compared to existing tools and generates no false positives, validating its effectiveness.

WANA's capability is limited to detecting blockchain-info dependency, with three false negatives in our dataset. This limitation arises from WANA's simplified modeling of on-chain database APIs, which substitutes on-chain data queries with random values of matching data types, resulting in missed execution branches.
For example, we identify a potential vulnerability in a contract where a sensitive function's execution depends on blockchain information as a call condition. 
In this case, as shown in Fig. \ref{fig:exm}, the code path contains a $\textsl{db\_get\_i64}$ function, whose internal operations WANA has not modeled, resulting in empty data retrieval. 
Since this path includes an assertion requiring non-empty data, WANA incorrectly concludes that the execution would be rolled back and the contract is secure, ultimately producing a false negative result.

% WASAI demonstrates lower precision and recall rates compared to our tool. 
% WASAI uses symbolic execution to guide the seed generation of fuzzy tests to improve code coverage.
% However, its use of symbolic execution is coarse-grained, especially for modeling on-chain data, which only focuses on data dependencies such as reads and writes, but not on internal operations.
% As a result, when confronted with vulnerabilities related to on-chain data, it is not possible to generate suitable seeds to cover the branch where the vulnerability is located, which ultimately leads to underreporting.
% Its accuracy is further compromised by oversimplified pattern design, such as flagging all calls to library functions quering blockchain info  (e.g., tapos\_block\_prefix and tapos\_block\_num) as vulnerability indicators.

% WASAI使用符号执行来指导模糊测试的种子生成，从而提高代码覆盖率。
% 然而，其使用的符号执行是粗粒度的，尤其是对链上数据的建模，只关注了读写等数据依赖，而不关注内部操作。
% 因此，在面对链上数据相关的漏洞时，无法生成合适的种子覆盖漏洞所在的分支，最终导致了漏报。

% Its reduced recall stems from coarse-grained symbolic execution in fuzzing tests, particularly for on-chain data dependencies. WASAI's imprecise simulation of data read-write relationships limits code coverage and hinders exploration of vulnerability paths dependent on specific on-chain database states. Its accuracy is further compromised by oversimplified pattern design, such as flagging all calls to library functions quering blockchain info  (e.g., tapos\_block\_prefix and tapos\_block\_num) as vulnerability indicators.

WASAI demonstrates lower precision and recall rates compared to our tool, primarily due to its limitations in on-chain API modeling and vulnerability pattern design. Although WASAI employs symbolic execution to guide fuzzy test seed generation for improved code coverage, its implementation is coarse-grained, particularly in modeling on-chain data where it only considers basic read and write dependencies while neglecting internal operations. This limitation impacts its effectiveness when analyzing vulnerabilities related to on-chain data, as it fails to generate suitable seeds to cover vulnerable code branches, resulting in missed detections. Additionally, WASAI's accuracy is further compromised by its oversimplified vulnerability pattern design, which incorrectly flags all blockchain information query functions (e.g., $\textsl{tapos\_block\_prefix}$ and $\textsl{tapos\_block\_num}$) as potential vulnerability indicators.
This lead WASAI to false positives in detecting \textit{Blockchain-info Dependency}.

We also compare the time taken by each tool to detect vulnerabilities. 
The results are shown in TABLE \ref{tab:timecomp}. 
Comparatively, WANA achieves the fastest detection times due to its one-round execution and simplified handling of on-chain data APIs. 
WASAI incurs higher overhead from multiple rounds of symbol execution-guided fuzzing,  and our tool's median detection time remains competitive with WASAI despite higher average detection times. 
Manual analysis of the ten most time-intensive contracts reveals that our tool's thorough analysis, though requiring up to an hour, successfully identifies vulnerabilities in cases where other tools produced false negatives by accessing their undetected branches.

\subsection{The experiments on real-world dataset}

To answer RQ2, we test \ourtool's effectiveness on the \textit{real-world dataset}. 
The evaluation reveals that 2,485 smart contracts contain vulnerabilities, distributed across four categories: 307 \textit{Missing Permission Check} vulnerabilities, 510 \textit{Rollback} vulnerabilities, 2,141 \textit{Integer Overflow} vulnerabilities, and 58 \textit{Blockchain-info dependency} vulnerabilities (with some contracts containing multiple vulnerabilities). 
The fact that these vulnerabilities affect over 48\% of the smart contracts in \textit{real-world dataset} underscores both their prevalence on the EOSIO platform and the urgent need for high-precision detection tools. 
Notably, \textit{Integer Overflow} emerges as the most common vulnerability, primarily due to the complexity introduced by multiple dependency libraries in WASM smart contracts, which challenges even experienced developers in manual detection efforts. 
This finding particularly emphasizes the importance of accurate automated detection tools in WASM smart contract development.

% Since there is no vulnerability label for the real-world contracts, we conduct a manual inspection of part of contracts from the real-world dataset to evaluate our tool's accuracy.
% As most of the real-world contracts are in WASM format, makes direct human readability challenging and costs human effort to inspect, we randomly choose 200 contracts.
% For each vulnerability type, we choose 50 contracts flagged as vulnerable and 50 flagged as non-vulnerable, using Ghidra for reverse engineering to facilitate code inspection. 
% Our inspection identifies 3 false-negative cases: 
% Two cases stemm from complex control flow structures that prevents our tool from exploring vulnerable code branches.
% And the other one case results from limitations in constructing concrete transactions.
% In the construction of transactions, we need to generate values for all the parameters by constraint solving.
% For complex types like nested structs, we simplify the process by randomly generate values for the elements of the struct, which may fail to achieve required on-chain database states for some branches. 
% Despite these limitations, our tool achieved results with 100\% precision and 97.09\% recall across the sample set, partly demonstrating its accuracy in real-world vulnerability detection. 

Since there is no vulnerability label for the real-world contracts, we conduct a manual inspection of selected contracts from the real-world dataset to evaluate our tool's accuracy. 
Due to the WASM format of most real-world contracts, which makes direct human readability challenging and requires significant effort to inspect, we randomly select 200 contracts for evaluation. 
Specifically, for each vulnerability type, we select 50 contracts flagged as vulnerable by \ourtool and 50 flagged as non-vulnerable, utilizing Ghidra~\cite{ghidra} for reverse engineering to facilitate manual inspection. 
Our inspection identifies 3 false-negative cases. 
Two cases stem from complex control flow structures that prevent our tool from exploring vulnerable code branches, while the third case results from limitations in constructing concrete transactions. 
In the transaction construction process, we need to generate values for all parameters through constraint solving. 
However, for complex types like nested structs, we simplify the process by randomly generating values for the struct elements, which may have failed to achieve the required on-chain database states for certain branches. 
Despite these limitations, our tool achieves results with 100\% precision and 97.09\% recall across the sample set, partly demonstrating its effectiveness in real-world vulnerability detection.

\subsection{The experiments evaluating our core components}
\label{subsec:componenttest}

\begin{table*}[]
\centering 
 \caption{A comparison of different version of \ourtool on \textit{vulnerability dataset}. }
 \label{tab:vercomp}
\begin{tabular}{c|c|c|c|c|c|c|c}
\hline
Components                                                                                               & Tool    & TP  & TN  & FP & FN & Timeout & Avg time \\ \hline
\multirow{3}{*}{\begin{tabular}[c]{@{}c@{}}On-chain API Emulation{[}$\times${]}\\ Concolic Execution{[}$\times${]}\\ Loop{[}$\times${]}\end{tabular}} & \multirow{3}{*}{\begin{tabular}[c]{@{}c@{}}\ourtool-con \\  \ourtool-sym \end{tabular}}   & \multirow{3}{*}{\begin{tabular}[c]{@{}c@{}}59\\ 96 \end{tabular}} & \multirow{3}{*}{\begin{tabular}[c]{@{}c@{}}321\\ 329\end{tabular}} & \multirow{3}{*}{\begin{tabular}[c]{@{}c@{}}17\\ 9 \end{tabular}} & \multirow{3}{*}{\begin{tabular}[c]{@{}c@{}}79\\ 42\end{tabular}} & \multirow{3}{*}{\begin{tabular}[c]{@{}c@{}}0\\ 31  \end{tabular}}     & \multirow{3}{*}{\begin{tabular}[c]{@{}c@{}}338.57s\\ 1520.12s  \end{tabular}} \\  
                                                                                               &        &   &  &   &  &       &    \\  
                                                                                               &        &   &  &   &  &        &    \\  \hline
\multirow{3}{*}{\begin{tabular}[c]{@{}c@{}}On-chain API Emulation{[}$\times${]}\\ Concolic Execution{[}$\times${]}\\ Loop{[}\checkmark{]}\end{tabular}} & \multirow{3}{*}{\begin{tabular}[c]{@{}c@{}}\ourtool-multicon \end{tabular}}   & \multirow{3}{*}{\begin{tabular}[c]{@{}c@{}}69 \end{tabular}} & \multirow{3}{*}{\begin{tabular}[c]{@{}c@{}}329\end{tabular}} & \multirow{3}{*}{\begin{tabular}[c]{@{}c@{}}9 \end{tabular}} & \multirow{3}{*}{\begin{tabular}[c]{@{}c@{}}69\end{tabular}} & \multirow{3}{*}{\begin{tabular}[c]{@{}c@{}}10  \end{tabular}}     & \multirow{3}{*}{\begin{tabular}[c]{@{}c@{}}474.36s  \end{tabular}} \\  
                                                                                               &        &   &  &   &  &       &    \\  
                                                                                               &        &   &  &   &  &        &    \\  \hline
\multirow{3}{*}{\begin{tabular}[c]{@{}c@{}}On-chain API Emulation{[}\checkmark{]}\\ Concolic Execution{[}$\times${]}\\ Loop{[}$\times${]}\end{tabular}} & \multirow{3}{*}{\begin{tabular}[c]{@{}c@{}}\ourtool-emu \end{tabular}}   & \multirow{3}{*}{\begin{tabular}[c]{@{}c@{}}105 \end{tabular}} & \multirow{3}{*}{\begin{tabular}[c]{@{}c@{}}338\end{tabular}} & \multirow{3}{*}{\begin{tabular}[c]{@{}c@{}}0 \end{tabular}} & \multirow{3}{*}{\begin{tabular}[c]{@{}c@{}}33\end{tabular}} & \multirow{3}{*}{\begin{tabular}[c]{@{}c@{}}15  \end{tabular}}     & \multirow{3}{*}{\begin{tabular}[c]{@{}c@{}}839.12s  \end{tabular}} \\  
                                                                                               &        &   &  &   &  &       &    \\  
                                                                                               &        &   &  &   &  &        &    \\  \hline
\multirow{3}{*}{\begin{tabular}[c]{@{}c@{}}On-chain API Emulation{[}\checkmark{]}\\ Concolic Execution{[}\checkmark{]}\\  Loop{[}$\times${]}\end{tabular}} & \multirow{3}{*}{\begin{tabular}[c]{@{}c@{}}\ourtool-single \end{tabular}}   & \multirow{3}{*}{\begin{tabular}[c]{@{}c@{}}104 \end{tabular}} & \multirow{3}{*}{\begin{tabular}[c]{@{}c@{}}338\end{tabular}} & \multirow{3}{*}{\begin{tabular}[c]{@{}c@{}}0 \end{tabular}} & \multirow{3}{*}{\begin{tabular}[c]{@{}c@{}}34 \end{tabular}} & \multirow{3}{*}{\begin{tabular}[c]{@{}c@{}}0  \end{tabular}}     & \multirow{3}{*}{\begin{tabular}[c]{@{}c@{}}321.58s  \end{tabular}} \\  
                                                                                               &        &   &  &   &  &       &    \\  
                                                                                               &        &   &  &   &  &        &    \\  \hline
\multirow{3}{*}{\begin{tabular}[c]{@{}c@{}}On-chain API Emulation{[}\checkmark{]}\\ Concolic Execution{[}\checkmark{]}\\  Loop{[}$\times${]}\end{tabular}} & \multirow{3}{*}{\begin{tabular}[c]{@{}c@{}}\ourtool-rnd \\  \ourtool \end{tabular}}   & \multirow{3}{*}{\begin{tabular}[c]{@{}c@{}}115\\ 120 \end{tabular}} & \multirow{3}{*}{\begin{tabular}[c]{@{}c@{}}338\\ 338\end{tabular}} & \multirow{3}{*}{\begin{tabular}[c]{@{}c@{}}0\\ 0 \end{tabular}} & \multirow{3}{*}{\begin{tabular}[c]{@{}c@{}}23\\ 18\end{tabular}} & \multirow{3}{*}{\begin{tabular}[c]{@{}c@{}}0\\ 0  \end{tabular}}     & \multirow{3}{*}{\begin{tabular}[c]{@{}c@{}}482.51s\\ 498.03s  \end{tabular}} \\  
                                                                                               &        &   &  &   &  &       &    \\  
                                                                                               &        &   &  &   &  &        &    \\  \hline

% \multirow{3}{*}{\begin{tabular}[c]{@{}c@{}}On-chain API Emulation{[}\checkmark{]}\\ Concolic Execution{[}\checkmark{]}\\  Loop{[}\checkmark{]}\end{tabular}} & \multirow{3}{*}{\begin{tabular}[c]{@{}c@{}}\ourtool \end{tabular}}   & \multirow{3}{*}{\begin{tabular}[c]{@{}c@{}}120 \end{tabular}} & \multirow{3}{*}{\begin{tabular}[c]{@{}c@{}}338\end{tabular}} & \multirow{3}{*}{\begin{tabular}[c]{@{}c@{}}0 \end{tabular}} & \multirow{3}{*}{\begin{tabular}[c]{@{}c@{}}18\end{tabular}} & \multirow{3}{*}{\begin{tabular}[c]{@{}c@{}}0  \end{tabular}}     & \multirow{3}{*}{\begin{tabular}[c]{@{}c@{}}498.03s  \end{tabular}} \\  
%                                                                                                &        &   &  &   &  &       &    \\  
%                                                                                                &        &   &  &   &  &        &    \\  \hline
% \begin{tabular}[c]{@{}c@{}}op {[}-{]} real{[}-{]} loop{[}-{]}\end{tabular}                 & \ourtool-emu       & 91  & 338 & 0  & 47 & 29      & 839.16s   \\ \hline
% \multirow{2}{*}{\begin{tabular}[c]{@{}c@{}}op{[}-{]}\\ real{[}-{]} loop{[}-{]}\end{tabular}} & \ourtool-single       & 104 & 338 & 0  & 34 & 0       & 321.58s    \\ \cline{2-8} 
%                                                                                    & \ourtool-rnd       & 115 & 338 & 0  & 23 & 0       & 482.51s    \\ \hline
% \begin{tabular}[c]{@{}c@{}}op {[}-{]} real{[}-{]} loop{[}-{]}\end{tabular}      & \ourtool & 120 & 338 & 0  & 18 & 0       & 498.03s    \\ \hline
\end{tabular}
\begin{tabular}{c}
[$\times$]: without the component\ \ \ \ [\checkmark]: with the component\\
Timeout: the number of contracts that cannot be analyzed in 1 hour    \\
\end{tabular}
\end{table*}

\textbf{Effectiveness of On-chain API Emulation.}
To answer RQ3, we first test the effectiveness of fine-grained emulation of on-chain APIs.
We implement \ourtool-con, which uses \ourtool's symbolic execution engine but doesn't emulate internal operations of on-chain APIs, instead using random concrete values to represent on-chain API return values.
Similarly, we implement \ourtool-sym, which represents on-chain API return values with symbolic values.
Correspondingly, we implement \ourtool-emu, which emulates internal operations of on-chain data.
Note that to separately evaluate the effectiveness of on-chain API emulation and other components, none of these three tools use concolic execution or coverage-guided loops.
Additionally, since \ourtool-con uses random return values, we run this tool three times and take the average of the results.

The results are illustrated in TABLE \ref{tab:vercomp}.
Our evaluation reveals that only \ourtool-emu avoids false positives.
This difference in accuracy directly stems from \ourtool-emu's ability to simulate internal operations of on-chain data APIs.
As mentioned in \ref{subsec:motivate}, without proper API operation modeling, both symbolic and random value approaches lose crucial constraint relationships between API parameters and internal operations, resulting in infeasible variable values and false positives.
On the other hand, \ourtool-emu also produces fewer false negatives.
Compared to \ourtool-con, since it uses random values to represent on-chain API return values, it might fail to cover certain execution branches, leading to false negatives.
While \ourtool-sym uses symbolic values to represent return values and won't miss execution branches, it ignores some relationships between on-chain data-related variables, introducing more symbolic variables to be solved in constraint solving, thus reducing analysis efficiency. This ultimately results in 31 contracts not completing analysis within 1 hour, therefore producing more false negatives than \ourtool-emu.
These results demonstrate that fine-grained emulation of on-chain APIs can improve the tool's analysis accuracy.

\begin{figure}
\centering
\includegraphics[scale=0.255]{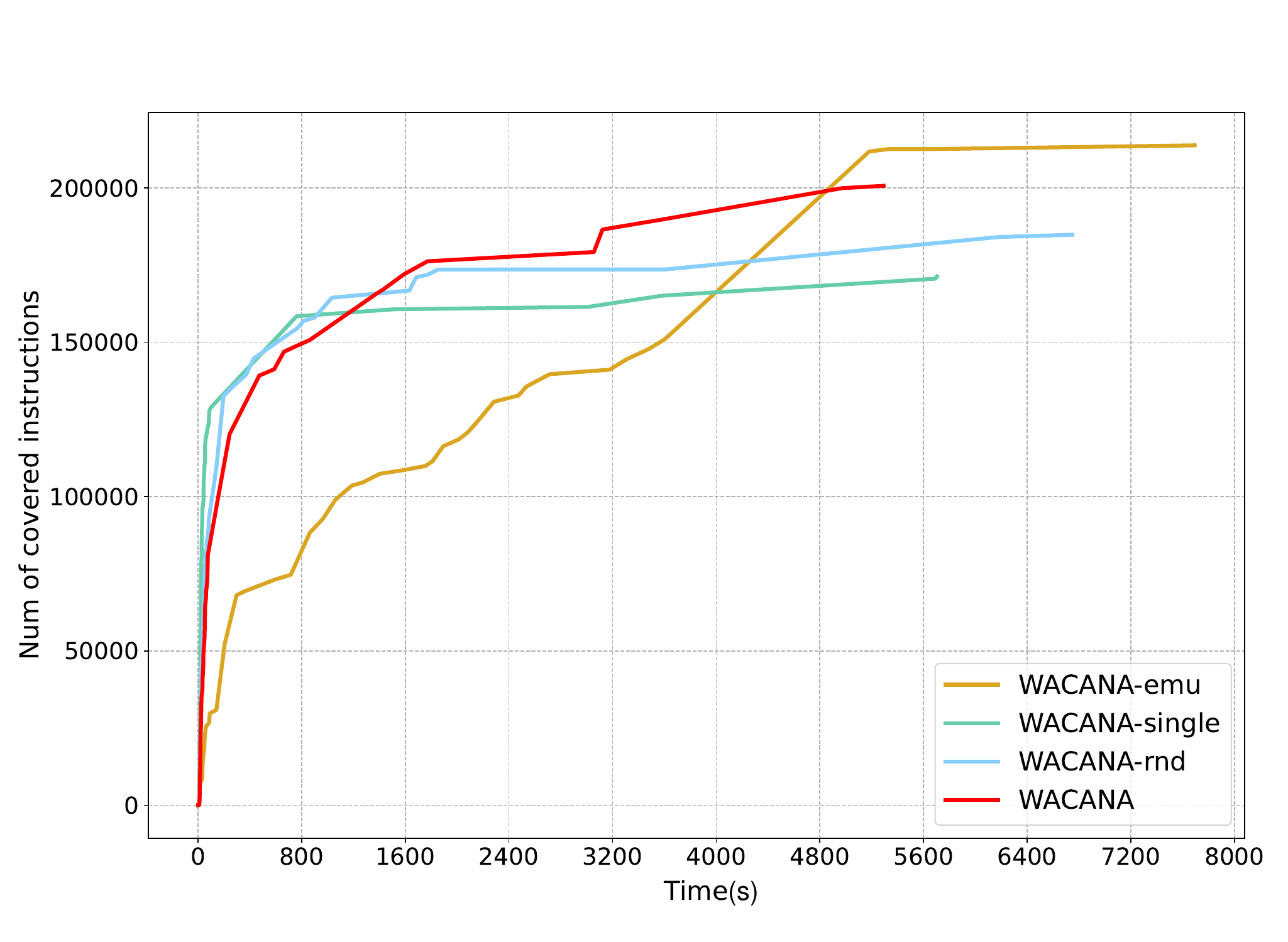} 
\caption{Growth in cumulative coverage of different tools with increasing analysis time.}
\label{fig:res}
\end{figure}

\textbf{Effectiveness of Concolic Execution.}
Next, we compare our tool with \ourtool-emu to illustrate the effectiveness of concolic execution.
% As discussed in Section \ref{subsec:motivate}, although emulating internal operations of on-chain APIs can increase analysis accuracy, it also brings additional time costs.
% For example, the average contract analysis time of \ourtool-emu is 2.5 times that of \ourtool-con.
% 如第x章所述，虽然模拟，但会带来更多的分析时间开销。
% 例如对比emu和con，可以发现emu的平均分析时间是con的2.5倍。
% As mentioned above, although \ourtool-emu emulates internal operations of on-chain APIs to achieve higher analysis accuracy than \ourtool-con, its average contract analysis time is also longer (2.5 times that of \ourtool-con).
% This is because \ourtool-emu uses symbolic values to represent on-chain data, which brings additional time costs to constraint solving and branch instruction execution.
% Comparatively, our tool, through multiple rounds of concolic execution, both accurately simulates internal operations of on-chain APIs while using concrete values to represent on-chain data to reduce time overhead.
Experimental results in TABLE \ref{tab:vercomp} show that \ourtool-emu's average analysis time is longer than our tool's (1.68 times), with 15 timeout cases occurring within the 1-hour time limit.
Further investigation reveals that these timeouts account for all 15 additional false negatives in \ourtool-emu's results compared to our tool.
Notably, extending the time limit to 7,200 seconds enables \ourtool-emu to complete analysis across all contracts, though it produces an identical number of false negatives as our tool.
This finding further validates the effectiveness of our concolic execution approach: it achieves significant efficiency improvements while maintaining detection accuracy across the vulnerability dataset.
Furthermore, we perform a coverage analysis between \ourtool-emu and our tool, whose results are illustrated in Fig. \ref{fig:res}.
Our tool reaches maximum coverage (200,651 instructions) at approximately 4,800 seconds, while \ourtool-emu requires 7,600 seconds to achieve its maximum of 213,755 instructions.
Although \ourtool-emu ultimately achieves slightly higher coverage, \ourtool's coverage growth rate significantly outperforms \ourtool-emu during the first 4,800 seconds.
This shows that \ourtool achieves a better balance between instruction coverage and analysis efficiency, an advantage brought by concolic execution.

\begin{figure}
\centering
\includegraphics[scale=0.55]{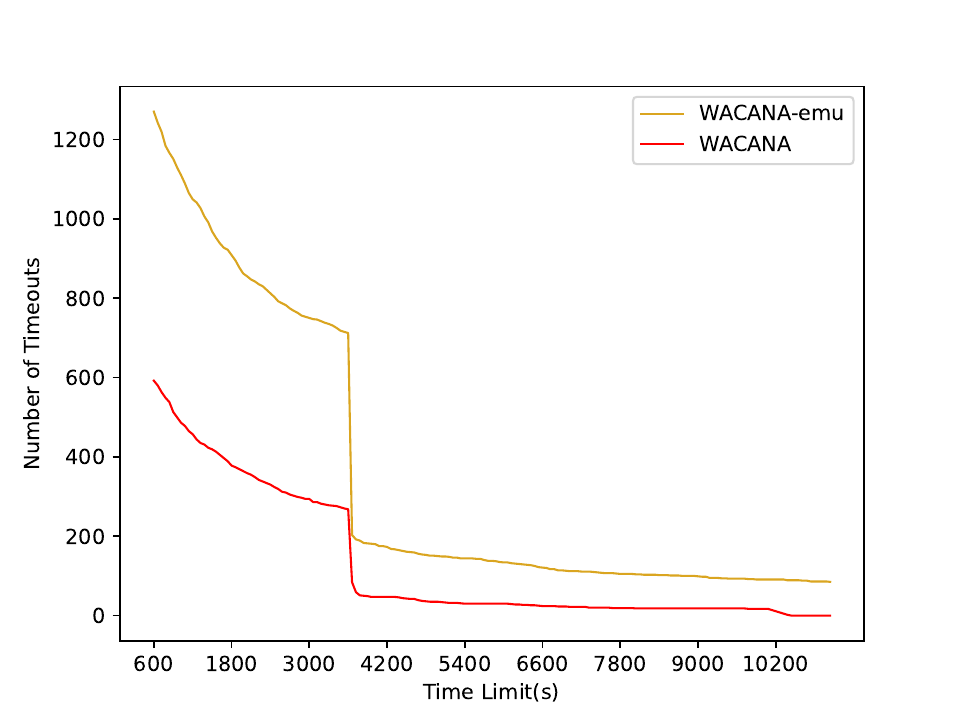} 
\caption{Number of timeout contracts for \ourtool and \ourtool-emu under different time limits.}
\label{fig:wildres}
\end{figure}

Additionally, we run both \ourtool-emu and our tool on the real-world dataset, with results shown in Fig. \ref{fig:wildres}.
With a time limit of 600 seconds (common time limit for other representative tools), \ourtool-emu fails to complete analysis for 1,270 contracts, accounting for 22.67\% of the entire dataset.
Our tool gets 592 timeouts, less than half of \ourtool-emu's timeout count.
When the time limit is set to 3,600 seconds (our tool's default time limit), \ourtool-emu has 712 timeouts, while our tool only has 268 timeouts.
Finally, when we extend the time limit to 10,800 seconds, we find that our tool can complete the analysis of all contracts, while \ourtool-emu still has 86 timeouts.

The above experimental results on the two datasets demonstrate that multiple rounds of concolic execution can balance analysis efficiency and accuracy.

\textbf{Effectiveness of Coverage-guided Loop.}
Lastly, we evaluate the effectiveness of the coverage-guided Loop.
For this, we implement \ourtool-rnd and \ourtool-single, which perform fine-grained modeling of on-chain data API internal operations and use concolic execution, but only perform single-round execution or use random strategies to guide the loop, respectively.
Since \ourtool-rnd uses a random strategy, we run this tool three times and take the average of the results.
Comparing the results of \ourtool-rnd, \ourtool-single, and our tool in TABLE \ref{tab:vercomp}, we can see that with no false positives, \ourtool-single has the most false negative results.
This is because all three tools concretize on-chain data through concolic execution, and under this premise, single-round execution leads to lower execution branch coverage, resulting in more false negatives.
While \ourtool-rnd, which guides multiple rounds of execution with a random strategy, has fewer false negatives compared to \ourtool-single.
However, it still gets more false negatives than our tool, indicating that coverage-guided strategies with clear objectives are superior to random strategies.

Similarly, we also explore how \ourtool-single and \ourtool-rnd's cumulative coverage increase over time, with results also shown in Fig. \ref{fig:res}.
Within 1,500 seconds, \ourtool-single and \ourtool-rnd's coverage growth is slightly faster than our coverage-guided strategy.
This is because \ourtool-single only executes one round, while \ourtool-rnd randomly selects functions to execute, thus saving some analysis time by not needing to calculate and compare coverage.
This can also be seen from the table where \ourtool-single and \ourtool-rnd's average analysis time is lower than our tool's.
However, as shown in the figure, after 1,500 seconds, our tool achieved higher instruction coverage than \ourtool-single and \ourtool-rnd, and reached the maximum value fastest.
In comparison, \ourtool-single and \ourtool-rnd's final maximum instruction coverage is both lower than our tool's and took a longer time, indicating that coverage-guided loops can more efficiently guide the tool to execute new branches.

Additionally, we implement \ourtool-multicon, which is similar to \ourtool-con but runs multiple rounds (set to 10) to improve analysis coverage.
The experimental results in TABLE \ref{tab:vercomp} show that combining multiple rounds of execution alone can improve accuracy and reduce false negatives and false positives.
However, simply using multiple rounds of execution cannot solve the limitation of lost on-chain data-related dependencies, still resulting in false positives, while also having execution branches that are difficult to cover within limited rounds, thus having more false negative results than our tool.
This experimental result demonstrates that using multiple rounds of execution alone cannot effectively improve analysis accuracy, and an effective combination of on-chain API and concolic execution in our tool is necessary.

\section{Threats to validity}

While our tool achieves better results than other representative tools in our experiments, it still has certain limitations:

\textbf{Dependence on Vulnerability Patterns.}
Similar to other representative tools, our tool uses predefined vulnerability patterns to detect contract vulnerabilities. As a result, it cannot automatically detect vulnerabilities arising from unknown patterns. 
However, our tool is designed with modularity, separating the vulnerability detector from other components, which makes it extensible. New types of vulnerabilities can be detected by manually summarizing their patterns and incorporating them into the vulnerability detector.

\textbf{Focus on Vulnerabilities Related to On-chain Data.}
Our tool primarily enhances the accuracy of detecting vulnerabilities related to on-chain data. For other types of vulnerabilities, such as fake EOS and fake receipts, our tool is not specifically optimized. Consequently, these vulnerabilities are not studied in our experiments. It is worth noting, though, that by integrating relevant patterns identified by tools like WANA, our tool could also cover these types of vulnerabilities.

\textbf{Omission Due to Concolic Execution.}
Because our tool concretizes on-chain data, some paths associated with specified on-chain data may not be covered by our tool. 
Although we implement a loop trying to maximize coverage, we still encounter some false negatives in our experiments. 
Notably, however, our tool ultimately reports only 18 false negatives, achieving higher recall than other representative tools.

\textbf{Inherent Challenges of Symbolic Execution.}
State space explosion~\cite{baldoni2018survey} is a common challenge for symbolic execution techniques. In our tool, we attempt to balance accuracy and analysis efficiency through concolic execution. Nonetheless, the state space explosion issue may still occur when our tool symbolically executes instructions unrelated to on-chain data.

\section{Related Work}
\label{sec:related}

\textbf{Static Analyzers for WASM Smart Contracts.}
Current static analyzers for WASM smart contracts primarily use symbolic execution and data flow analysis.
EOSAFE is a static analysis framework that employs WASM bytecode-based symbolic execution, with vulnerability-specific heuristic pruning and semantic-aware library function simulation.
WANA extends this approach by improving scalability and providing broader vulnerability coverage across both Ethereum and EOSIO platforms.
EOSIOAnalyzer~\cite{li2022eosioanalyzer}, converts WASM code into Register Transfer Language (RTL) to enable data flow analysis-based vulnerability detection. Meanwhile, EOSDFA~\cite{li2022poster} implements a data flow analysis method using Octopus~\cite{octopus}, a security analysis framework for WASM contracts that supports pointer access operations through static single assignment (SSA) transformation.
EVulHunter~\cite{quan2019evulhunter} also utilizes Octopus for constructing control flow graphs and simulates the execution of specific basic blocks to detect pattern-based vulnerabilities.
VETEOS~\cite{veteos} is an analyzer designed to vet “Groundhog Day”, a kind of vulnerabilities that other tools cannot detect, by defining it as a control and data dependency problem.
Different from the above analyzers, EXGEN~\cite{jin2022exgen} is an exploit generation framework for Ethereum and EOSIO contracts, which can generate attack contracts aimed at vulnerable contracts and verify the exploitability of vulnerabilities in them.

\textbf{Dynamic Analyzers for WASM Smart Contracts.}
Existing dynamic analyzers for WASM smart contracts mainly use fuzz testing-based approaches.
EOSFUZZER generates fuzz test inputs based on smart contract interface information, executes contracts using a modified WASM virtual machine, and performs vulnerability detection based on the collected execution data.
Building on EOSFUZZER, WASAI adopts a hybrid method that combines fuzz testing with symbolic execution. It constructs path constraints and performs mutation and constraint solving to generate adaptive seeds, improving upon EOSFUZZER's random seed generation strategy and increasing path coverage.
Unlike EOSFUZZER, which modifies the EOSIO WASM virtual machine, WASAI achieves better portability by using contract instrumentation to collect execution logs.

\textbf{Analysis Methods for WASM Web Applications.}
There are also some researches on WASM web application security.
For instance, Weikang et al.~\cite{bian2019poster} develop WASM command execution path modeling to detect cryptojacking attacks, while Daniel et al.~\cite{lehmann2020everything} investigate WASM binary vulnerabilities and proposed mitigation strategies.
Jonathan et al.~\cite{protzenko2019formally} advance formal verification techniques for WASM cryptographic implementations, providing verified encryption components.
Furthermore, Wasabi~\cite{lehmann2019wasabi} establish a comprehensive framework for WASM dynamic analysis, enabling features such as instruction counting, call graph extraction, memory access tracking, and taint analysis through JavaScript-based binary instrumentation.

\section{Conclusion}

We propose and implement \ourtool, a concolic analyzer for WASM smart contracts that enhances the detection accuracy of vulnerabilities related to on-chain data while optimizing analysis efficiency through integration of concrete execution and coverage-guided loop.
The experimental results on a vulnerability dataset show that \ourtool outperforms other tools in terms of accuracy and the components of \ourtool, e.g., emulation of on-chain data APIs, are necessary.
Additionally, an evaluation on 5,602 real-world contracts demonstrates \ourtool's practical effectiveness.

\bibliographystyle{ACM-Reference-Format}
\bibliography{main}

\end{document}